\documentclass[10pt,twocolumn,letterpaper]{article}
\usepackage{iccv}
\usepackage{times}
\usepackage{epsfig}
\usepackage{graphicx}
\usepackage{amsmath}
\usepackage{amssymb}
\usepackage{commands}
\usepackage{tikz}
\usepackage{multicol}
\usepackage{multirow}
\usepackage{color}
\usepackage{xcolor}
\usepackage{colortbl,booktabs}
\newcommand*\circled[1]{\tikz[baseline=(char.base)]{
            \node[shape=circle,draw,inner sep=0.2pt] (char) {#1};}}
\newcolumntype{g}{>{\columncolor[gray]{.9}}c}

\usepackage[pagebackref=true,breaklinks=true,letterpaper=true,colorlinks,bookmarks=false]{hyperref}

\iccvfinalcopy 

\def\iccvPaperID{10232} 

\ificcvfinal\fi

\title{Physics-Driven Turbulence Image Restoration with Stochastic Refinement}

\author{%
  Ajay Jaiswal$^{1*}$, Xingguang Zhang$^{2*}$, Stanley H. Chan$^{2}$, Zhangyang Wang$^{1}$ \\
  $^{1}$University of Texas at Austin, $^{2}$Purdue University\\
  \texttt{\{ajayjaiswal, atlaswang\}@utexas.edu} \texttt{\{zhan3275, stanchan\}@purdue.edu}\\
 }

\begin{document}
\maketitle
\def \thefootnote{*} \footnotetext{Equal contribution.}\def \thefootnote{\arabic{footnote}}
\begin{abstract}
    Image distortion by atmospheric turbulence is a stochastic degradation, which is a critical problem in long-range optical imaging systems. A number of research has been conducted during the past decades, including model-based and emerging deep-learning solutions with the help of synthetic data. Although fast and physics-grounded simulation tools have been introduced to help the deep-learning models adapt to real-world turbulence conditions recently, the training of such models only relies on the synthetic data and ground truth pairs. This paper proposes the Physics-integrated Restoration Network (PiRN) to bring the physics-based simulator directly into the training process to help the network to disentangle the stochasticity from the degradation and the underlying image. Furthermore, to overcome the ``average effect" introduced by deterministic models and the domain gap between the synthetic and real-world degradation, we further introduce PiRN with Stochastic Refinement (PiRN-SR) to boost its perceptual quality. Overall, our PiRN and PiRN-SR improve the generalization to real-world unknown turbulence conditions and provide a state-of-the-art restoration in both pixel-wise accuracy and perceptual quality. Our codes are available at \url{https://github.com/VITA-Group/PiRN}.
\end{abstract}

\section{Introduction} \label{sec:intro}
Atmospheric turbulence (AT) is one of the major sources of degradation in 
long-range passive imaging systems. It arises due to random spatiotemporal fluctuations in the index of refraction \cite{Fried_1966_a, Tatarski_1967_a}. When accumulating over the distance, turbulence oftentimes leads to degraded image quality with random pixel displacement and blurring \cite{Chan_2022_a}.  Such degradation is very common in settings when the object distance is long, and the exposure is short \cite{Tatarski_1967_a}, causing a substantial drop in the performance of long-range passive imaging and its downstream tasks, such as human recognition, detection, and tracking. Developing restoration methods to mitigate the turbulence effect is important. However, anisoplanatic turbulence has two major properties: \circled{1} the geometric warping and blur are entangled with each other; \circled{2} the point spread function is spatially and temporally varying, which makes the problem harder than other image restoration problems.

Turbulence mitigation algorithms have been studied for several decades by the image processing community. Since capturing real-world corrupted and clean image pairs is almost impossible, those conventional methods are model-based \cite{Delbracio_2015_a, Gilles_2016_a, Huebner_2012_a, Oreifej_2013_a, Anantrasirichai_2013_a, Droege_2012_a, Hardie_2017_b, Milanfar_2013_a, Mao_2020_a}. Recently, data-driven approaches have been introduced with synthetic data from various turbulence simulators. Existing deep learning works explored both deterministic and stochastic methods to solve the turbulence mitigation problem. In the deterministic approaches \cite{Mao_2022_a, Zhang_2022_a, Anantrasirichai_2022_a, Jin_2021_a, Lau_2021_b, Nieuwenhuizen_2019_a, Nair_2021_a, Hoffmire_2021_a, rai2022removing, feng2023turbugan}, a turbulence mitigation network is trained on a synthetic dataset to minimize pixel-level distortion between the output and ground truth images. The generalization capability of those works is fully limited by the training images. Moreover, the deterministic models learn to ``fill up" a probability space that covers all possible clean images, which leads to unnatural output with an average effect. Although adversarial training \cite{Jin_2021_a, Lau_2021_b, rai2022removing} has been used to alleviate this problem, it could make the model more vulnerable to small perturbation on the input \cite{Choi_2019_ICCV, choi2022deep}. The stochastic approaches \cite{nair2023ddpm, chung2022parallel} could produce more natural images with certain robustness. However, they are more likely to hallucinate as they do not have a physics-grounded degradation model, and the generative models they used are unconditional diffusion models trained on general datasets. The large domain gap between the testing images and the training set distribution and the lack of a good forward model as a connection oftentimes cause the output of generative models unreliable.

\begin{figure*}[ht]
    \centering
    \includegraphics[width=\linewidth]{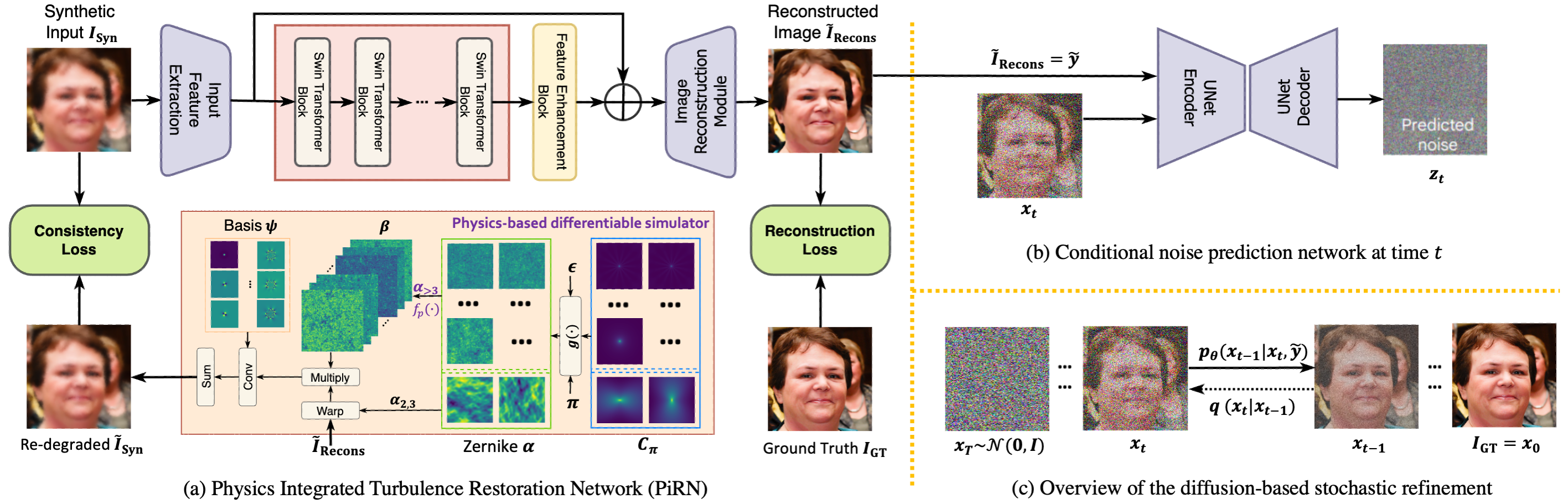}
    \caption{The overview of our PiRN with stochastic refinement model. The PiRN (a) training utilizes a physics-based differentiable simulator to map the stored image back to the input to enforce the consistency between the forward and restoration processes. We use the intermediate restoration from PiRN as a reliable condition of a diffusion model at each denoising step (b). We perform 10-20 stochastic iterations (c) to sample a natural image with high perceptual quality from the well-constraint image space.}
    \vspace{-0.4cm}
    \label{fig:main_architecture}
\end{figure*}

In this work, we propose a two-stage method to improve both the fidelity \cite{naeem2020reliable} and perceptual quality \cite{whang2022deblurring} for single-frame turbulence mitigation. To improve \underline{fidelity}, we propose to tightly couple a physics-based turbulence simulator \cite{mao_p2s, dfp2s} in the training paradigm of our turbulence restoration backbone. Specifically, we re-degrade the reconstructed image and align it with the original input to enforce the consistency between image formation and restoration. Our consistency enforcement could facilitate the separation of image semantics and turbulence profiles by injecting the turbulence conditions in the training loop. With the physics-integrated restoration network (PiRN), we experimentally found the above strategy significantly improves the generalization across multiple real-world datasets with varying turbulence strength.

 To improve \underline{perceptual quality}, unlike \cite{Jin_2021_a, Lau_2021_b, rai2022removing}, which used the adversarial training, we build a stochastic posterior sampler by training a conditional denoising diffusion probabilistic model (DDPM) \cite{ho2020denoising} that takes the output of the PiRN as a reliable condition and generates high-quality natural images within the constraint sample space. Our experiments found direct conditioning on the degraded images could destabilize diffusion training due to the failure to capture the complicated degradation. We take advantage of the deterministic reconstructor PiRN and use its restored image as a more constrained condition. Our divide-and-conquer strategy suggests PiRN focus on handling turbulence strength variations and facilitating \textit{stochastic refinement (SR)} to mitigate the gap between the rough restoration and real-world image distribution.

Our overall framework PiRN-SR enjoys the benefit of high fidelity and perceptual quality while adapting to a wide range of physical attributes of turbulence degradation (i.e., generalization). PiRN-SR combines PiRN with stochastic refinement in a plug-and-play fashion. It iteratively performs 10-20 denoising steps to significantly boost \textit{perceptual quality and robustness to additional perturbation} without compromising the fidelity. Our primary contributions can be briefly summarized as follows:

\begin{itemize}
    \item [$\square$] We show how a fully differentiable physics-based simulator can be tightly coupled with the DL restoration paradigm. We improve the generalization to multiple physical attributes of turbulence (distance to object, camera settings, etc.) with the help of our diligently curated synthetic data generation strategy.

    \item [$\square$] Our proposed framework PiRN-SR demonstrates how a carefully trained conditional diffusion model can be used as a plug-and-play stochastic refiner to generate high perceptual quality results from turbulence-degraded input images at marginal inference overhead.

    \item [$\square$] Extensive experiments and ablation across synthetic and multiple real-world popular turbulence benchmarks demonstrate our method achieves state-of-the-art reconstruction quality in both pixel-wise accuracy and perceptual quality. 
\end{itemize}


\section{Method}
\subsection{Zernike-based turbulence forward model} \label{sec:simulator}
The forward model of the degradation caused by atmospheric turbulence to an image $\boldsymbol{I}$ is \cite{Hirsh_2010_a}:
\begin{equation}
    \boldsymbol{O} = \boldsymbol{h}_{\vx}\circledast\boldsymbol{I}+\boldsymbol{n},
    \label{eq: eq1}
\end{equation}
where $\boldsymbol{I}\in \mathbb{R}^{H\times W}$ is the clean input image, $\boldsymbol{O}\in \mathbb{R}^{H\times W}$ is the captured image degraded by turbulence, $\circledast$ is the convolution, $\vx$ is the 2-D spatial position of pixels and $\boldsymbol{n}\in \mathbb{R}^{H\times W}$ is the Gaussian random noise. Since the point spread function (PSF) $\boldsymbol{h}_\vx$ varies with $\vx$, the degradation is \emph{spatially varying}.
According to \cite{Chan_2022_a}, $\vh_\vx$ may be written as a function of pixel-shifting $\boldsymbol{\mathcal{T}}$ (closely associated with ``tilt'' in the optical literature) and blur $\boldsymbol{\mathcal{B}}_{\vx}$ in the strict order:
\begin{equation}
    \boldsymbol{O} \approx \boldsymbol{\mathcal{B}}_{\vx}\circledast \boldsymbol{\mathcal{T}}(\boldsymbol{I})+\boldsymbol{n}.
    \label{eq: forward_process}
\end{equation}

The recent Zernike-based turbulence simulators \cite{chimitt2020simulating, dfp2s} model the degradation as a wide sense stationary (WSS) random field of the phase distortion, which is represented by Zernike polynomials $\{\mZ_i\}$ \cite{Noll_1976_a} as a basis, with coefficients $\va_{\vx, i}$, where  $i\! \in \! \{1,2,3,\cdots,36\}$. Given a set of camera and atmospheric \emph{protocol} $\boldsymbol{\pi}$, the autocorrelation $\boldsymbol{C}_{\boldsymbol{\pi}}$ of the Zernike random field can be drawn by this simulator. From $\boldsymbol{C}_{\boldsymbol{\pi}}$ and Gaussian white noise $\boldsymbol{\epsilon} \in \mathbb{R}^{H \times W \times 36}$, the WSS Zernike random field $\va_{\vx}$ can be generated via Fourier Transform. We denote this function by $\va_{\vx} = g(\boldsymbol{C}_{\boldsymbol{\pi}}, \boldsymbol{\pi}, \boldsymbol{\epsilon})$

Among all 36 coefficients, $i=1$ denotes the current component, $i={2,3}$ controls the $\boldsymbol{\mathcal{T}}$ by a constant scale, and the rest high order Zernike coefficients contribute to the blur effect. The phase distortion corresponding to high-order Zernike coefficients can be efficiently translated to 100 PSFs basis $\boldsymbol{\psi}$ and coefficients $\boldsymbol{\beta}_{\vx}$ via the phase-to-space (P2S) \cite{mao_p2s} transform $\boldsymbol{\beta}_{\vx} = f_{p}(\va_{\vx, \{i \geq 4\}})$. The spatially varying blur kernel can be finally written as:
\begin{equation}
    \boldsymbol{\mathcal{B}}_{\vx} \approx \sum_{k=1}^{100} \boldsymbol{\beta}_{\vx,k} \boldsymbol{\psi}_k = \sum_{k=1}^{100} f_{p}(\va_{\vx, \{i \geq 4\}})_{k} \boldsymbol{\psi}_k.
    \label{eq: p2s_psf_basis}
\end{equation}
For the detailed expression of $g(\cdot)$ and $f_{p}(\cdot)$, we refer readers to read section I of the supplementary material. 

\subsection{Physics-integrated restoration network (PiRN)}
Since $g$ can be implemented by spectrum decomposition, $f_p$ is a small neural network \cite{mao_p2s}, and $\boldsymbol{\psi}$ are fixed basis kernels, the forward model conditioned on the turbulence protocol $\boldsymbol{\pi}$ and random seed $\boldsymbol{\epsilon}$ is differentiable. This property suggests the forward process can be embedded into the training loop and effectively provide the turbulence prior through gradients, facilitating the invariance of the reconstruction towards the stochasticity of the degradation.

The degradation profile controlled by $\boldsymbol{\epsilon}$ is different for each frame, which makes anisoplanatic turbulence get spatially and temporally varying distortions. Conventional CNN-based general image restoration methods applying the fixed kernels on all spatial locations may not be adequate to solve the location adaptive problem \cite{Mao_2022_a}. Motivated by the \textit{necessity of input-adaptive and location-adaptive filtering effect}, our physics-integrated restoration network (PiRN) uses a transformer-based network to capture and recover the spatially- and instance-varying turbulence effects. 

Overall, PiRN architecture is composed of a Swin-based deep backbone module, a convolution-based image reconstruction module, and a physics-based differentiable forward model, as described in detail in the following.

\label{sec:simulator integration}
\subsubsection{Phase-to-Space differentiable forward process}
The primary contribution to PiRN design is the integration of the Phase-to-Space differentiable turbulence simulator in the training paradigm. Conventional turbulence mitigation networks \cite{Mao_2022_a, Zhang_2022_a, Jin_2021_a, Anantrasirichai_2022_a, Nair_2021_a} trained their models only with the low-quality and reference high-quality pairs, the generalization capability of their networks only relies on their synthetic method and training data, hence inept at adapting to varying turbulence protocols and profiles. \cite{Mao_2022_a, Li_2021_a, feng2023turbugan} proposed to re-map the restored image to the degraded image. However, the re-mapping of \cite{Mao_2022_a} is based on an empirical design without a clear physical meaning. \cite{Li_2021_a, feng2023turbugan} are multi-frame methods, their reconstruction requires minutes or hours of refinement, and the adaptation to real-world cases is yet highly limited.

In PiRN, we propose to integrate the well-established physics of turbulence described in section \ref{sec:simulator} and explore its experimental benefits on unseen turbulence protocols in both synthetic and real-world datasets.
More specifically, we store the degradation protocol $\boldsymbol{\pi}$ and random seed $\boldsymbol{\epsilon}$ along with the degraded image $\mathbf{I}_\text{Syn}$ in the synthetic data generation stage. During training, the degraded image is first restored by the PiRN backbone and reconstruction module, the restored image $\mathbf{\widetilde{I}}_{\text{Recons}}$ is then passed through the simulator with $\boldsymbol{\pi}$ and $\boldsymbol{\epsilon}$ to re-degrade the original input to $\widetilde{\mathbf{I}}_\text{Syn}$. Precisely, the function of this module $\mathbf{SIM(\cdot)}$ can be summarized as:

\begin{equation}
    \underset{\substack{\text{Reconstructed Clean PiRN Output} \\ \text{and Data Synthesis Protocol}}}{\underbrace{\{\mathbf{\widetilde{I}}_\text{Recons}, \boldsymbol{\pi}, \boldsymbol{\epsilon}\}}}
    \rightarrow \mathbf{SIM(\cdot)} \rightarrow 
    \underset{\substack{\text{Re-degraded} \\ \text{Output Image}}}{\underbrace{\widetilde{\mathbf{I}}_\text{Syn}}}
\end{equation}

We force $\mathbf{I}_\text{Syn}$ and $\widetilde{\mathbf{I}}_\text{Syn}$ to be aligned using the $\ell_1$ loss. Since this remapping enforces the restoration to be consistent with the degradation process, we call it \emph{consistency loss}. During training, by gradient descent, the turbulence prior from the simulator can be injected into the network and help enhance the invariance of the turbulence effect. Despite its simplicity, we found it significantly improves the adaptability across multiple real-world datasets with varying turbulence strength.

\subsubsection{Swin-based deep feature backbone} Swin Transformer \cite{liu2021swin} has recently shown great success in modeling long-range dependency with shifted window schemes. Although it has been explored for many image restoration works \cite{liang2021swinir, Zamir_2022_CVPR, Wang_2022_CVPR, chen2022simple}, its potential for turbulence mitigation which requires spatially varying operations is still unexplored. As shown in Figure \ref{fig:main_architecture}, the deep feature backbone of PiRN architecture is a sequence of Swin Transformer blocks (RSTB). The RSTB utilizes several Swin Transformer layers for local attention and cross-window interaction. Finally, for feature enhancement,  we add a convolution layer at the end of RSTB blocks for feature enhancement and use a residual connection to provide a shortcut for feature aggregation \cite{liang2021swinir}. 
Before RSTB, the input feature extraction uses convolution layers to extract shallow features. Those features preserve low-frequency information of the image, induce convolutional inductive bias in the early stage, and improve the representation learning capability of transformer blocks \cite{xiao2021early}. The details of the architecture are provided in the supplementary material.

\subsubsection{Image Reconstruction Network} Our upsample image reconstruction module restores the high-quality image by decoding the deep features generated by the Swin-based backbone with reference to shallow features from input feature extraction. With the long skip connection, our network can transmit the low-frequency information directly to the reconstruction network, which can help the deep feature extraction module focus on high-frequency information and stabilize training. Our Image reconstruction module is a sequence of convolutional layers with LeakyReLU that projects enriched features back to low dimension feature map corresponding to the reconstructed clean image $\mathbf{\tilde{I}}_\text{Recons}$. Precisely, the role of the reconstruction module $\mathbf{REC(\cdot)}$ can be summarized as:
\begin{equation}
    \underset{\substack{\text{Deep and Shallow Features of} \\ \text{Input Image $\mathbf{I}_\text{Syn}$}}}{\underbrace{\{\mathbf{D_{I_\text{Syn}}, S_{I_\text{Syn}}}\}}}
    \rightarrow \mathbf{REC(\cdot)} \rightarrow 
    \underset{\substack{\text{Reconstructed Clean} \\ \text{Output Image}}}{\underbrace{\widetilde{\mathbf{I}}_\text{Recons}}}
\end{equation}

PiRN optimization requires the joint optimization of reconstruction loss with ground truth and the consistency loss as shown in Figure \ref{fig:main_architecture}. We formulate the loss as follows:
\begin{equation}
    \mathcal{L}_{total} = \alpha \cdot ||\mathbf{I}_\text{GT}, \widetilde{\mathbf{I}}_\text{Recons}||_{1}
    + (1 - \alpha) \cdot ||\mathbf{I}_\text{Syn}, \widetilde{\mathbf{I}}_\text{Syn}||_{1}
    \label{eq:total_loss}
\end{equation}

\subsection{Diffusion-based stochastic refinement}
Although the turbulence simulation tool is physics-grounded, the domain gap still exists between synthetic and real-world turbulence. Besides, the restored images from the PiRN also suffer from the problem of deterministic methods. In order to overcome this and make the output more natural and closer to the target dataset distribution, we use Denoising Diffusion Probabilistic Models (DDPM) \cite{sohl2015deep, ho2020denoising, song2021scorebased} as a stochastic sampler. DDPMs are generative models with a Markov chain that constructs samples iteratively from a joint distribution:
\begin{equation}
    p_{\theta}(\vx_{0:T}) = p_{\theta}^{(T)}(\vx_{T})\prod_{t=0}^{T-1}p_{\theta}^{(t)}(\vx_t|\vx_{t+1})
    \label{eq: diffusion_backward}
\end{equation}
Where $\vx_{T}$ is random Gaussian noise and $\vx_{0}$ is the ground truth. In our framework, the high-quality image space follows a posterior distribution conditioned on the pre-restored image $\widetilde{\vy}=\widetilde{\mathbf{I}}_\text{Recons}$ from our deterministic reconstructor:
\begin{equation}
    p_{\theta}(\vx_{0:T}|\widetilde{\vy}) = p_{\theta}^{(T)}(\vx_{T})\prod_{t=0}^{T-1}p_{\theta}^{(t)}(\vx_t|\vx_{t+1}, \widetilde{\vy})
    \label{eq: diffusion_condition_backward}
\end{equation}
 This conditional diffusion sampler could have an unconditional variational inference distribution:
\begin{equation}
    q(\vx_{1:T}|\vx_{0}) = q^{(T)}(\vx_{T}|\vx_{0})\prod_{t=1}^{T-1}q^{(t)}(\vx_t|\vx_{t+1}, \vx_0)
    \label{eq: diffusion_forward}
\end{equation}
Whose conditional transition distribution of the forward and backward diffusion process can be modeled by Gaussian parameterization. For $p_{\theta}^{(t)}$ and $q^{(t)}$, we reduce the evidence lower bound (ELBO) objective by solving the following noise prediction loss \cite{ho2020denoising}:
\begin{equation}
    \mathcal{L} = \sum_{t=1}^{T-1} \gamma_t \mathbb{E}_{(\vx_0, \vz_t, \widetilde{\vy})} \left [  ||\vz_t-\vz_{\theta}(\vx_t,\widetilde{\vy}, t) ||^{2} \right ]
    \label{eq: diffusion_loss}
\end{equation}
Where $\vz_t$ is the noise added to the forward diffusion process at time $t$, $\vx_{t}=\sqrt{\overline{\alpha}_{t}}\vx_{0}+\sqrt{1-\overline{\alpha}_t}\vz_{t}$ is the noised clean image at denoising step $t$, and $\vz_{\theta}(\vx_t,\widetilde{\vy}, t)$ is a network with learnable parameters to predict the noise $\vz_t$. $\overline{\alpha}_{t}$ is a fixed scalar to control the diffusion schedule  \cite{ho2020denoising}.

We perform gradient descent to train $\vz_{\theta}(\vx_t,\widetilde{\vy}, t)$ like \cite{saharia2022image}. When $\vz_{\theta}(\vx_t,\widetilde{\vy}, t)$ is trained, the restoration can be converted to a diffusion posterior sampler $\hat{\vx}_{0}\sim p_{\theta}(\vx_{0}|\widetilde{\vy})$ defined in Eq \ref{eq: diffusion_condition_backward}. As shown in Figure \ref{fig:main_architecture}, along with the ground truth $\mathbf{I}_\text{GT}$, we pair one restored image from PiRN. Our experiments found direct conditioning on the degraded images ($\mathbf{I}_\text{Syn}$) could destabilize diffusion training because the turbulence profiles are random, and the diffusion model can't capture the complicated degradation. During inference with PiRN-SR, we iteratively perform 10-20 denoising steps with our diffusion sampler upon the restoration output of PiRN to boost its perceptual quality.  

\begin{figure}[ht]
    \centering
    \includegraphics[width=\linewidth]{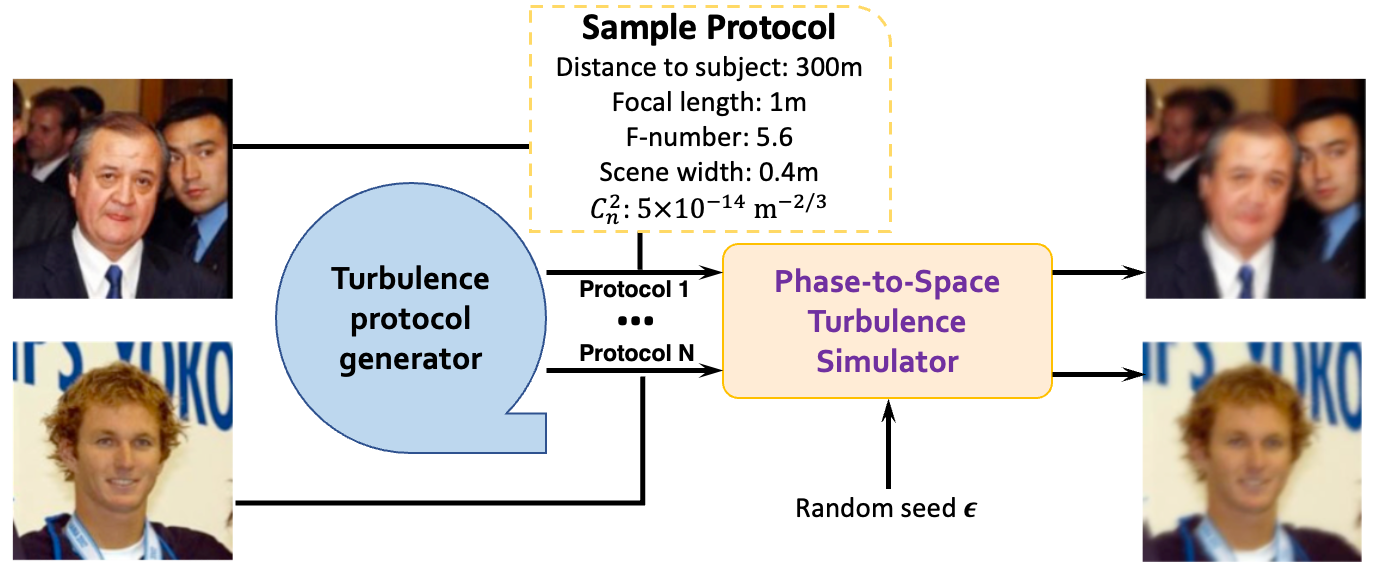}
    \caption{Overview of our synthetic data (PiRN-Syn) simulation. }
    \label{fig:my_label}
    \vspace{-0.5cm}
\end{figure}

\subsection{PiRN-Syn: Synthetic Data Generation Strategy} \label{sec:data_generation}
Because of the scarcity of real clean-distortion image pairs, data-driven approaches have to rely on turbulence simulators for synthetic data. Despite there have been some efforts in turbulence simulation in the image processing community \cite{lau, nair, Lachinova_2007_a, Mao_2022_a}, they highly overlook the fact that real-world turbulence profile is affected by the aperture and focal length of the imaging system, distance to the object, field of view, wavelength, and other environmental conditions (temperature, humidity, wind speed, and so on). This ignorance restricted the generalization capabilities of those methods for unknown real-world degradation. 

Following section \ref{sec:simulator}, we provide an \textit{easy-to-follow synthetic data generation strategy} capturing a wide variety of camera parameters and atmospheric conditions (represented by the measurable indicators $\mathcal{C}_n^2$) using the improved Zernike-based simulator \cite{chimitt2020simulating, dfp2s}. Our synthetic data generation strategy (Table \ref{table:synthesis_parameters}) has been curated based on the analysis to cover the potential short-exposure turbulence profiles observed in around 1000+ hours of video footage of approximately 1,000 subjects within different environmental conditions and camera settings \cite{cornett2023expanding}. When setting the parameters, we first select the distance and field of view, then the focal length and f-number ranges could be determined based on real-world camera models. We choose the $C_{n}^{2}$ range to set the turbulence effect to be neither too strong nor weak. PiRN-Syn consists of 100,000 degraded images for training and 50,000 degraded images for testing, which are generated using different synthesis protocols from Table \ref{table:synthesis_parameters} with 2000 and 1000 unique instances from \cite{places_dataset}. 

In addition, to study how our new turbulence mitigation algorithm performs under different conditions, we classify the turbulence strength into multiple levels: \emph{weak, medium, strong}. The details of how we set the threshold is provided in the supplementary material.

\begin{table}
\centering
\resizebox{\linewidth}{!}{
\begin{tabular}{cccccc}
\hline
Distance (m) & Focal length (m) & F-number & Scene width (m) & $C_{n}^{2} (10^{-14}\times \text{m}^{-2/3})$\\
\hline
 \multirow{2}{*}{[200, 400]} & \multirow{2}{*}{[1, 2]} & $\{8, 11\}$ & [0.2, 0.5] & [3, 7] \\\cline{3-5}
      &     & $\{5.6, 8, 11\}$  & [0.5, 1] & [6, 30] \\\cline{2-5}
 \multirow{2}{*}{[400, 600]} & \multirow{2}{*}{[1, 2.5]} & $\{8, 11, 16\}$ & [0.4, 0.8] & [2, 6] \\\cline{3-5}
                &      &  $\{5.6, 8, 11\}$  & [0.8, 1.5] & [6, 30] \\\cline{2-5}
 \multirow{2}{*}{[600, 800]} & \multirow{2}{*}{[1, 3]} & $\{11, 16\}$ & [0.5, 1.2] & [2, 5] \\\cline{3-5}
                &    &  $\{8, 11\}$  & [1.2, 2] & [5, 30] \\
\hline
\end{tabular}}
\vspace{0.1cm}
\caption[The detailed parameter ranges.]{Parameter range, where $[a,b]$ means uniform sampling from continuous range (a, b), and $\{\}$ indicates uniform sampling from a discrete set, all rows are chosen with identical probability.}
\vspace{-0.4cm}
\label{table:synthesis_parameters}
\end{table}

\section{Experimental Setup}
\subsection{Datasets and Training Setup}
To train the PiRN network, we utilized the PiRN-Syn synthetic dataset, which is generated by our curated data generation strategy as discussed in Section \ref{sec:data_generation}. The dataset comprises 100,000 degraded images for training and 50,000 degraded images for testing, generated using various synthesis protocols from Table \ref{table:synthesis_parameters}, with 2000 and 1000 unique instances from \cite{places_dataset}, respectively. The diffusion network is trained using high-quality images from \cite{karras2019style}, paired with the restored output of its degraded version from PiRN. For training PiRN, we used a learning rate of $1e-4$ and employed the cosine annealing scheduler to gradually decrease the learning rate over 100,000 iterations. The training of the diffusion network closely followed the optimal settings in \cite{saharia2022image}. In the beginning, during the first 5000 epochs of PiRN training, we set the scaling factor $\alpha$ in Equation \ref{eq:total_loss} to 1 to provide an easy optimization landscape for PiRN, and then set it to 0.9 for the remaining training iterations. More implementation details can be found in the supplementary.

\begin{figure*}[ht]
    \centering
    \includegraphics[width=\linewidth,trim={12em 8em 10em 8em},clip]{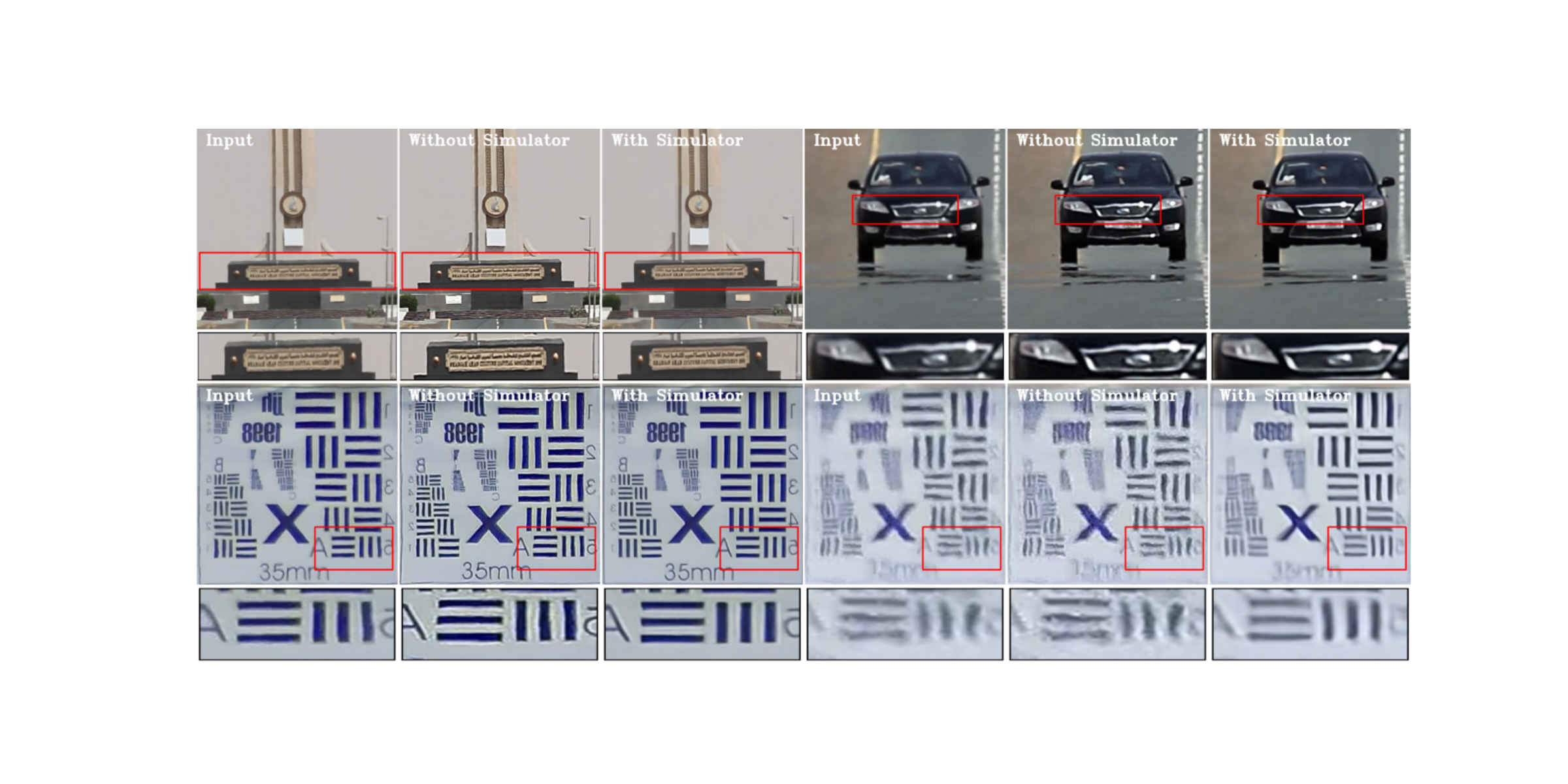}
    \caption{Qualitative performance of PiRN with and without our simulator integration on real-world turbulence benchmark CLEAR (Row1) and OTIS (Row2). Note that both models are trained using exactly the same PiRN-Sync train dataset for a fair comparison. Results in row 2 represent the ability of PiRN to generalize to different turbulence strengths.}
    \label{fig:with_without_simulator}
    \vspace{-0.4cm}
\end{figure*}

\begin{figure*}
    \centering
    \includegraphics[width=\linewidth,trim={14em 18em 10em 18em},clip]{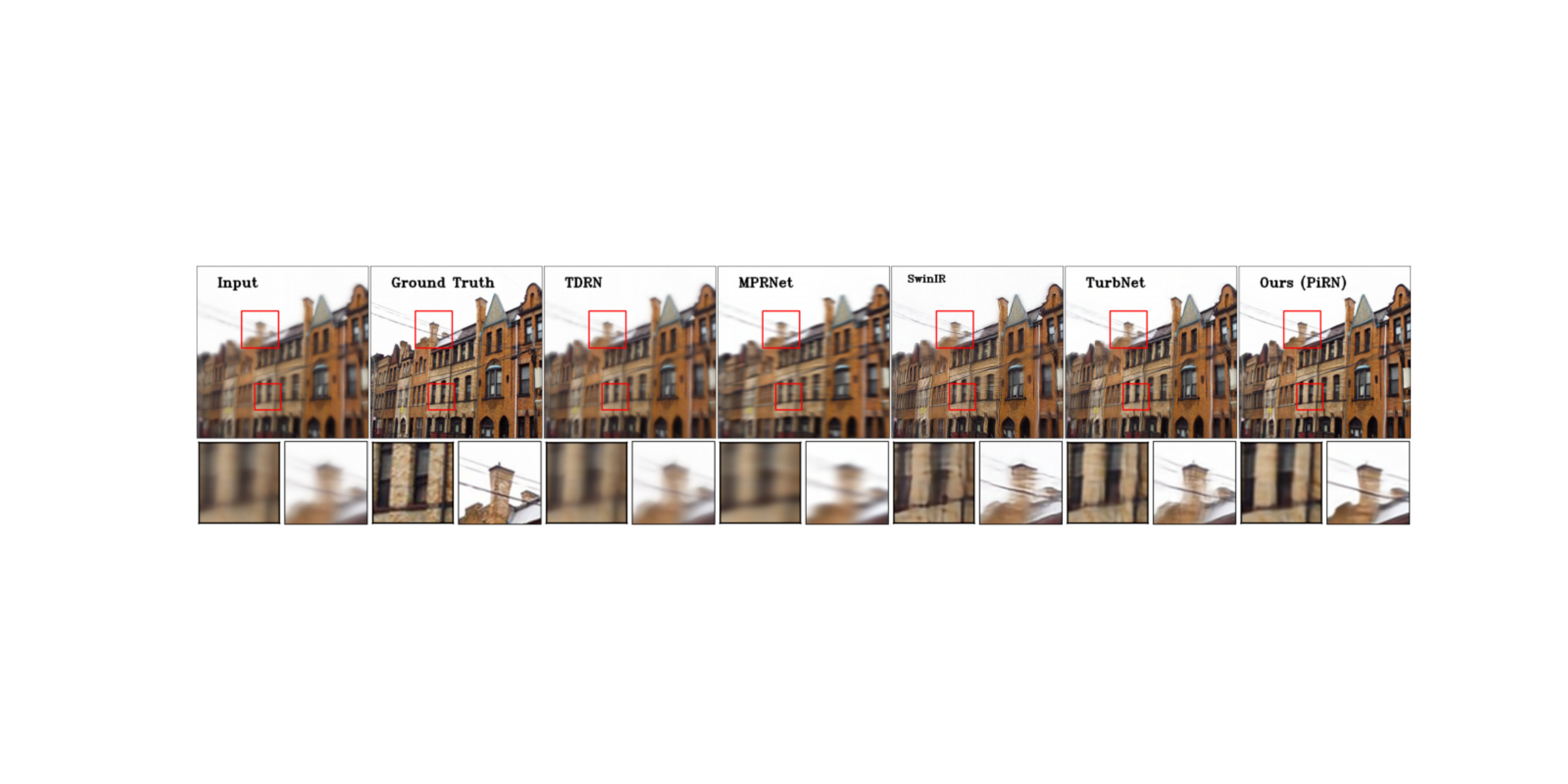}
    \caption{Qualitative performance comparison of our proposed method on synthetic PiRN-Syn (test) dataset wrt. SOTA baselines.}
    \label{fig:synthetic_data_comparison}
    \vspace{-0.4cm}
\end{figure*}

\subsection{Evaluation Protocol}
Our proposed framework incorporates a strong backbone, differentiable turbulence forward model, and a diffusion posterior sampler to achieve high-quality restoration. In order to validate the effectiveness of our design, we have designed experiments to answer several key questions:
\vspace{-0.4cm} 
\paragraph{\textbf{RQ1:}} How does the simulator-integrated PiRN design perform and generalize compared to classical state-of-the-art (SOTA) restoration architectures in both synthetic and real-world settings?

\vspace{-0.4cm}
\paragraph{\textbf{RQ2:}} Does simulator-in-loop training enhance the restoration adaptability of the proposed framework to varying levels of turbulence strength?

\vspace{-0.4cm}
\paragraph{\textbf{RQ3:}} How effective is posterior diffusion sampling in improving the perceptual quality of the restored images without impacting the standard PiRN metrics (PSNR, SSIM)?

\vspace{-0.4cm}
\paragraph{\textbf{RQ4:}} How effective is PiRN/PiRN-SR for ad-hoc downstream applications (e.g., detection and recognition)?

\vspace{0.1cm}
For our experiments, we have used our PiRN synthetic test set, along with a variety of real-world turbulence benchmark datasets, including OTIS \cite{Gilles_2017_a}, CLEAR\cite{Anantrasirichai_2022_a}, Heat Chamber and Turbulence Text Data \cite{Mao_2022_a}. We trained all baseline methods using identical settings and datasets, relying on their official GitHub implementation to ensure a fair comparison. Further details regarding our evaluation datasets can be found in the supplementary material.

\begin{figure*}
    \centering
    \includegraphics[width=0.9\linewidth,trim={14em 17em 10em 17em},clip]{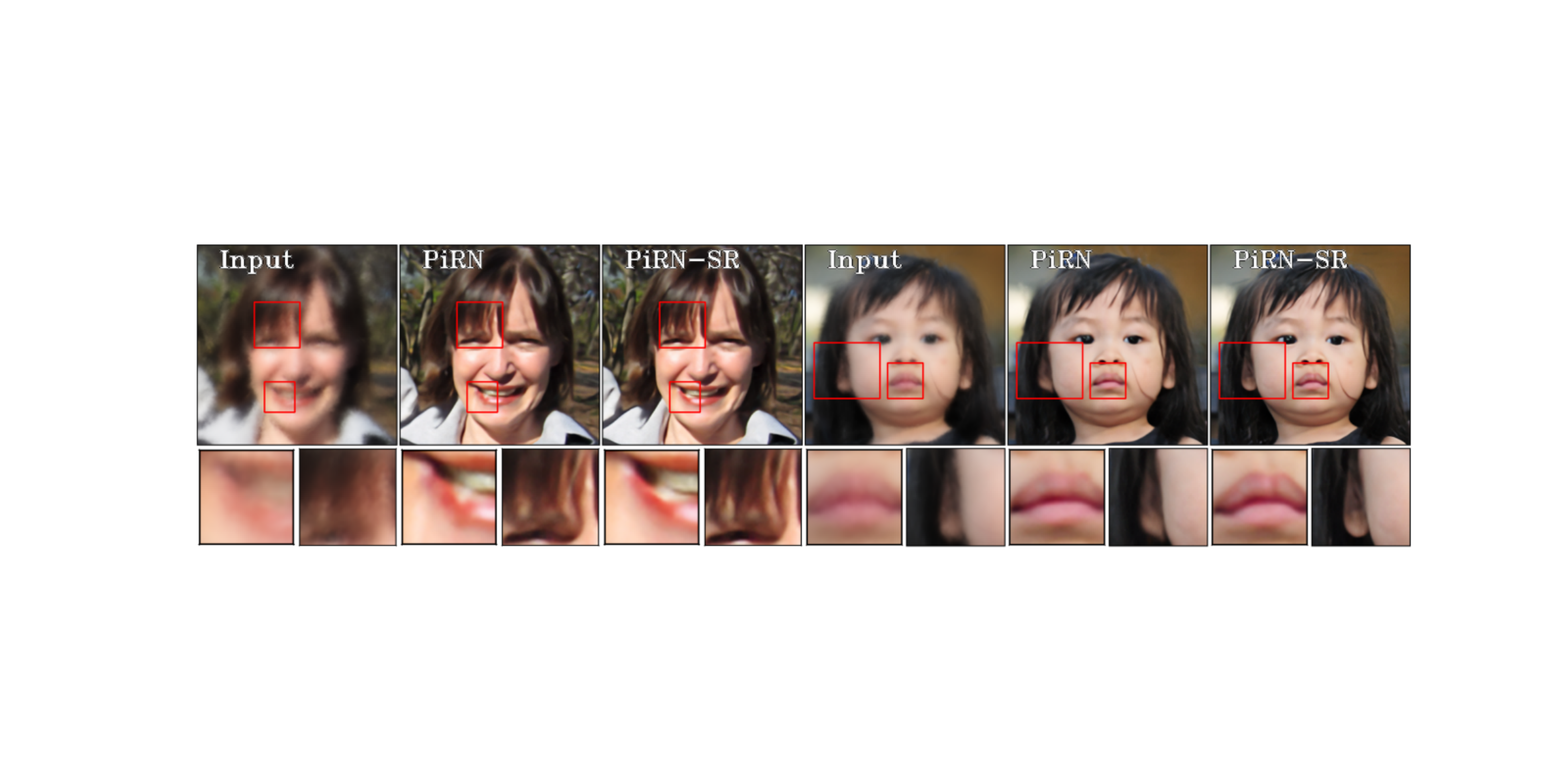}
    \caption{Qualitative performance benefits of using posterior diffusion sampling in improving the perceptual quality of the restored image. Our stochastic refinement module can be observed minimizing the gap between the rough restoration and real-world image distribution. }
    \vspace{-0.4cm}
    \label{fig:with_without_diffusion}
\end{figure*}

\begin{figure*}
    \centering
    \includegraphics[width=0.9\linewidth,trim={14em 15em 10em 13em},clip]{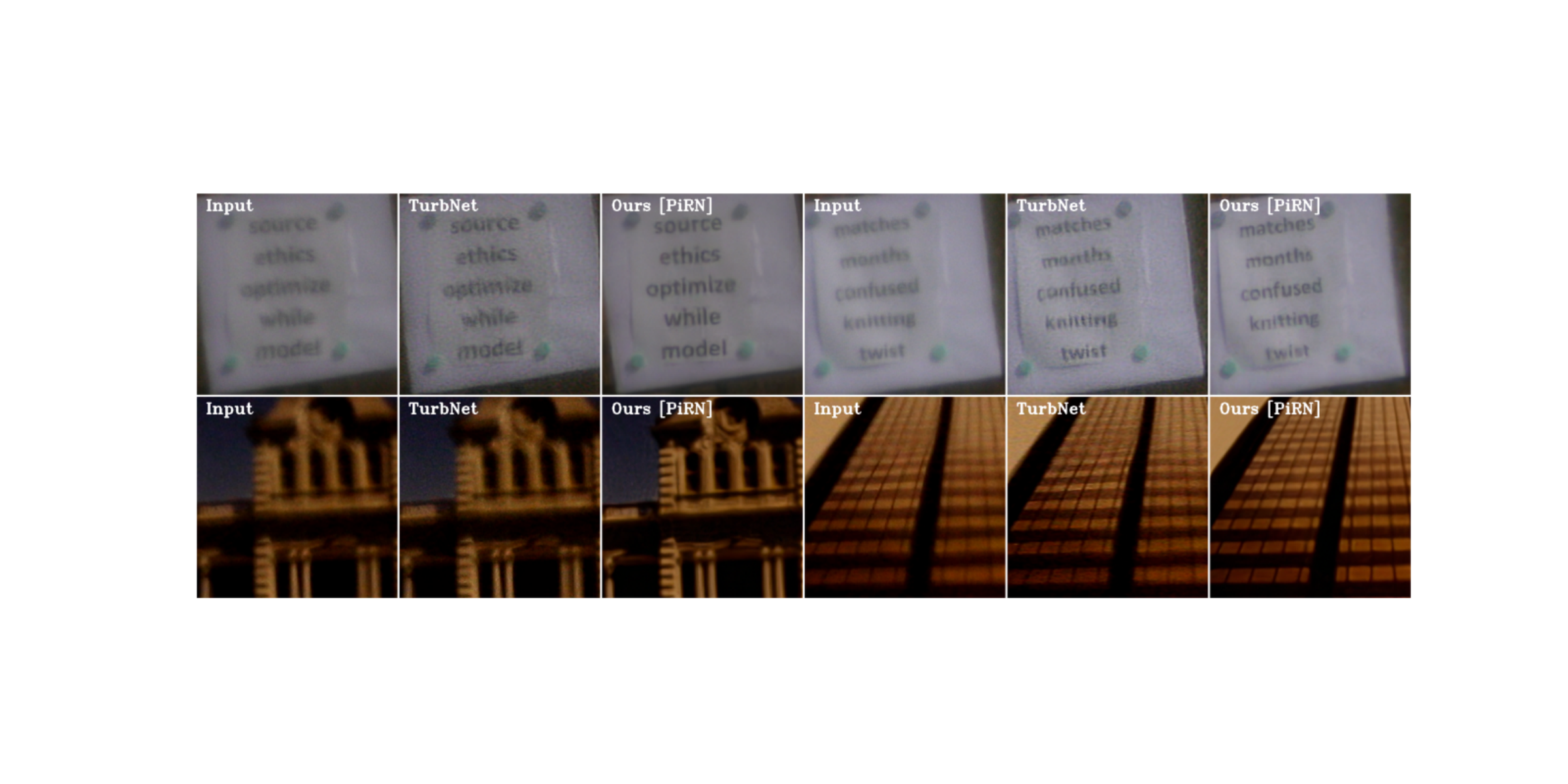}
    \caption{Qualitative performance of our proposed PiRN network on real-world Turbulence Text and Heat Chamber \cite{Mao_2020_a} datasets.}
    \vspace{-0.4cm}
    \label{fig:real_data_comparison}
\end{figure*}

\subsection{Simulator Integrated PiRN design and state-of-the-art restoration methods}
In this section, our main focus is to address research questions RQ1 and RQ2 by evaluating the advantages of integrating the turbulence forward model into the restoration training process to align with the synthetic data generation. To answer RQ1, which aims to determine the necessity of PiRN design in terms of performance and generalization, we conducted a comparative study by assessing the performance of PiRN with several state-of-the-art classical restoration methods. The results, presented in Table \ref{tab:rq1_syn_real_generalization}, demonstrate significant benefits of PiRN design on our synthetic test set and a real-world turbulence benchmark dataset Heat Chamber, in comparison to other baselines. It is noteworthy that our design outperforms general image restoration models significantly and also achieves higher performance than the recent turbulence-based design TurbNet \cite{Mao_2022_a}, with a margin of $+1.68$ (PSNR) and $+0.78$ (PSNR) on our synthetic test set and Heat Chamber dataset, respectively. Moreover, Figure \ref{fig:with_without_simulator} depicts the qualitative performance of PiRN with and without our simulator integration on real-world turbulence benchmark CLEAR and OTIS, trained on the same PiRN-Sync dataset. It is evident that our design plays a significant role in modeling the turbulence degradation operator and generalizes very well on the real-world unseen degradation.

\begin{table}[ht]
\centering
\setlength{\aboverulesep}{0pt}
\setlength{\belowrulesep}{0pt}
\resizebox{0.95\linewidth}{!}{\begin{tabular}{lcccc}
\hline
\multirow{2}{*}{Method} & \multicolumn{2}{c}{\texttt{PiRN-Syn} (Test)} & \multicolumn{2}{c}{\texttt{Heat Chamber}} \\ \cmidrule{2-5}
& PSNR $(\uparrow)$ & SSIM $(\uparrow)$ & PSNR $(\uparrow)$ & SSIM $(\uparrow)$ \\
\hline
TDRN \cite{Yasarla2021ICIP} & 19.48 & 0.5288 & 18.42 & 0.6424\\
MPRNet \cite{mehri2021mprnet} & 21.93 & 0.5819 & 18.12 & 0.6379\\
Uformer \cite{wang2022uformer} & 22.20 & 0.6133 & 18.68 & 0.6577\\
Restormer \cite{zamir2022restormer} & 22.45 & 0.6274 & 19.12 & 0.6840\\
SwinIR \cite{liang2021swinir} & 22.67 & 0.6301 & 19.43 & 0.6901\\
TurbNet \cite{Mao_2022_a}  & 23.72 & 0.6749 & 19.76 & 0.6934\\
\hline
\rowcolor[gray]{0.9} 
Ours [PiRN] & \underline{25.40} & \underline{0.7198} & \underline{20.54} & \underline{0.7102} \\
\rowcolor[gray]{0.9} 
Ours [PiRN-SR] & \textbf{25.61} & \textbf{0.7204} & \textbf{20.59} &  \textbf{0.7115}\\
\hline
\end{tabular}}
\vspace{0.2cm}
\caption{Performance comparison of our proposed method wrt. SOTA classical methods on our synthetic test set and real-world Heah Chamber\cite{Mao_2022_a} dataset.}
\vspace{-0.4cm}
\label{tab:rq1_syn_real_generalization}
\end{table}

Furthermore, we investigate the generalization capability of PiRN for different turbulence strengths (RQ2). Table \ref{tab:RQ2_strength_evaluation} presents the results of our evaluation, which confirm the ability of our proposed simulator-integrated design (PiRN) to handle different levels of turbulence strength smoothly in our synthetic test set. PiRN consistently outperforms TurbNet by a margin of $+0.81, +0.77, $ and $+1.12$ (PSNR), with significant benefits observed for strong degradation strength. The qualitative evaluation of the turbulence strength adaptation ability of PiRN on the real-world OTIS benchmark is also presented in Figure \ref{fig:with_without_simulator} (Row 2), which is consistent with our synthetic quantitative evaluation.

\begin{table}[ht]
\centering
\setlength{\aboverulesep}{0pt}
\setlength{\belowrulesep}{0pt}
\resizebox{1\linewidth}{!}{\begin{tabular}{lcccccc}
\hline
\multirow{2}{*}{Method} & \multicolumn{2}{c}{Weak} & \multicolumn{2}{c}{Medium} & \multicolumn{2}{c}{Strong}\\ \cmidrule{2-7}
& PSNR & SSIM & PSNR & SSIM & PSNR & SSIM \\
\hline
TDRN \cite{Yasarla2021ICIP} & 26.61 & 0.703 & 23.44 & 0.659 & 21.64 & 0.591\\
MPRNet \cite{mehri2021mprnet} & 27.01 & 0.716 & 24.76 & 0.672 & 22.98 & 0.620\\
Uformer \cite{wang2022uformer} & 27.76 & 0.721 & 25.30 & 0.687 & 23.54 & 0.633 \\
Restormer \cite{zamir2022restormer} & 27.91 & 0.729& 25.68 & 0.689 & 23.56 & 0.638 \\
SwinIR \cite{liang2021swinir} & 28.02 & 0.736& 25.96 & 0.691 & 23.87 & 0.645\\
TurbNet \cite{Mao_2022_a}  & 28.31& 0.759 & 26.07 & 0.738 & 23.90 & 0.649\\
\hline
\rowcolor[gray]{0.9} 
Ours [PiRN] & \underline{29.12} & \underline{0.763} & \underline{26.84} & \underline{0.747} & \underline{25.02} & \underline{0.662}\\
\rowcolor[gray]{0.9} 
Ours [PiRN-SR] & \textbf{29.15} & \textbf{0.766} & \textbf{26.86} & \textbf{0.747} & \textbf{25.07} & \textbf{0.665}\\
\hline
\end{tabular}}
\vspace{0.1cm}
\caption{Performance comparison of our proposed method across different turbulence strength wrt. other baselines.}
\vspace{-0.4cm}
\label{tab:RQ2_strength_evaluation}
\end{table}

\begin{table}[h!]
\centering
\small
\resizebox{\linewidth}{!}{\begin{tabular}{lccc|g}
\hline
 & Raw Input &  SwinIR\cite{liang2021swinir} & TurbNet\cite{Mao_2022_a} & \textbf{Ours [PiRN]}\\
\hline
AWDR$(\uparrow)$ & 0.623 &  0.740 & 0.758 & \textbf{0.776}\\
AD-LCS$(\uparrow)$ & 5.076 & 7.002 & 7.314 & \textbf{7.498} \\
\hline
\end{tabular}}
\vspace{0.1cm}
\caption{Performance comparison of state-of-art restoration baselines with respect to TurbNet on the Turbulence Text Dataset.}
\vspace{-0.4cm}
\label{table:ocr_challenege}
\end{table} 

\begin{figure*}[ht]
    \centering
    \vspace{-0.2cm}
    \includegraphics[width=0.9\linewidth]{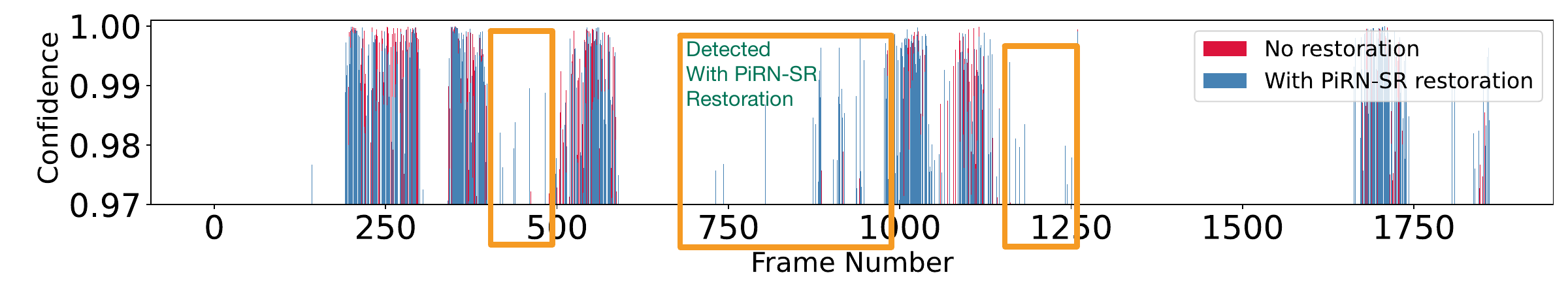}
    \caption{Performance of ad-hoc MTCNN-based face detector \cite{zhang2016joint} on a real-world video captured under atmospheric turbulence (camera distance 400m) \cite{cornett2023expanding}  wrt. its restored version from PiRN-SR. Each line indicates if a person's face has been detected in that frame with a given confidence score. It can be observed that PiRN-SR restoration significantly helps in improving the performance of MTCNN.}
    \vspace{-0.4cm}
    \label{fig:restoration_detection}
\end{figure*}

\vspace{-0.05cm}
\subsection{The importance of stochastic refinement}
\vspace{-0.15cm}
In this section, we investigate the effectiveness of posterior diffusion sampling in improving the perceptual quality of PiRN outputs (RQ3). Figure \ref{fig:with_without_diffusion} visually demonstrates the benefits of recursively performing 15 denoising steps using the DDIM sampler on the restoration output of PiRN. It is evident that these inexpensive refinement iterations can considerably improve the perceptual quality of the PiRN outputs, making them more natural. To further quantify the perceptual improvements, we utilize popular perceptual metrics such as NIQE, NRQM, and LPIPS. Table \ref{table:syn_nriq_nqrm_lipis} presents the performance of PiRN-SR compared to PiRN and other high-performing baselines. Our results indicate that integrating the diffusion sampler with the PiRN network significantly enhances the perceptual performance across all the above metrics on our PiRN-Syn test set compared to the state-of-the-art TurbNet baseline. Importantly, these benefits can be obtained in a plug-and-play fashion, depending on resource availability, without compromising the standard pixel-wise performance metrics such as PSNR and SSIM (Table \ref{tab:rq1_syn_real_generalization}). Specifically, PiRN-SR shows an improvement of $\sim 1.00$ (NIQE), $\sim 0.643$ (NRQM), and $\sim 0.023$ (LPIPIS) over PiRN with only marginal computational cost.

\begin{table}[ht]
\centering
\small
\resizebox{\linewidth}{!}{\begin{tabular}{l|cc|gg}
\hline
 &  SwinIR\cite{liang2021swinir} &  TurbNet\cite{Mao_2022_a} & PiRN & PiRN-SR\\
\hline
NIQE $(\downarrow)$ & 7.4914 & 7.0163 & \underline{6.5852} & \textbf{5.5847}\\
NRQM  $(\uparrow)$& 3.6423 & 3.7790 & \underline{3.9810} & \textbf{4.6239}\\
LPIPIS  $(\downarrow)$& 0.4585 & 0.4369  & \underline{0.4204} & \textbf{0.3978}\\
\hline
\end{tabular}}
\vspace{0.1cm}
\caption{Quantative perceptual performance benefits of PiRN-SR wrt. PiRN and other baselines on PiRN-Syn (test) dataset.}
\vspace{-0.4cm}
\label{table:syn_nriq_nqrm_lipis}
\end{table} 

\begin{figure*}
    \centering
    \includegraphics[width=0.9\linewidth,trim={14em 17em 10em 17em},clip]{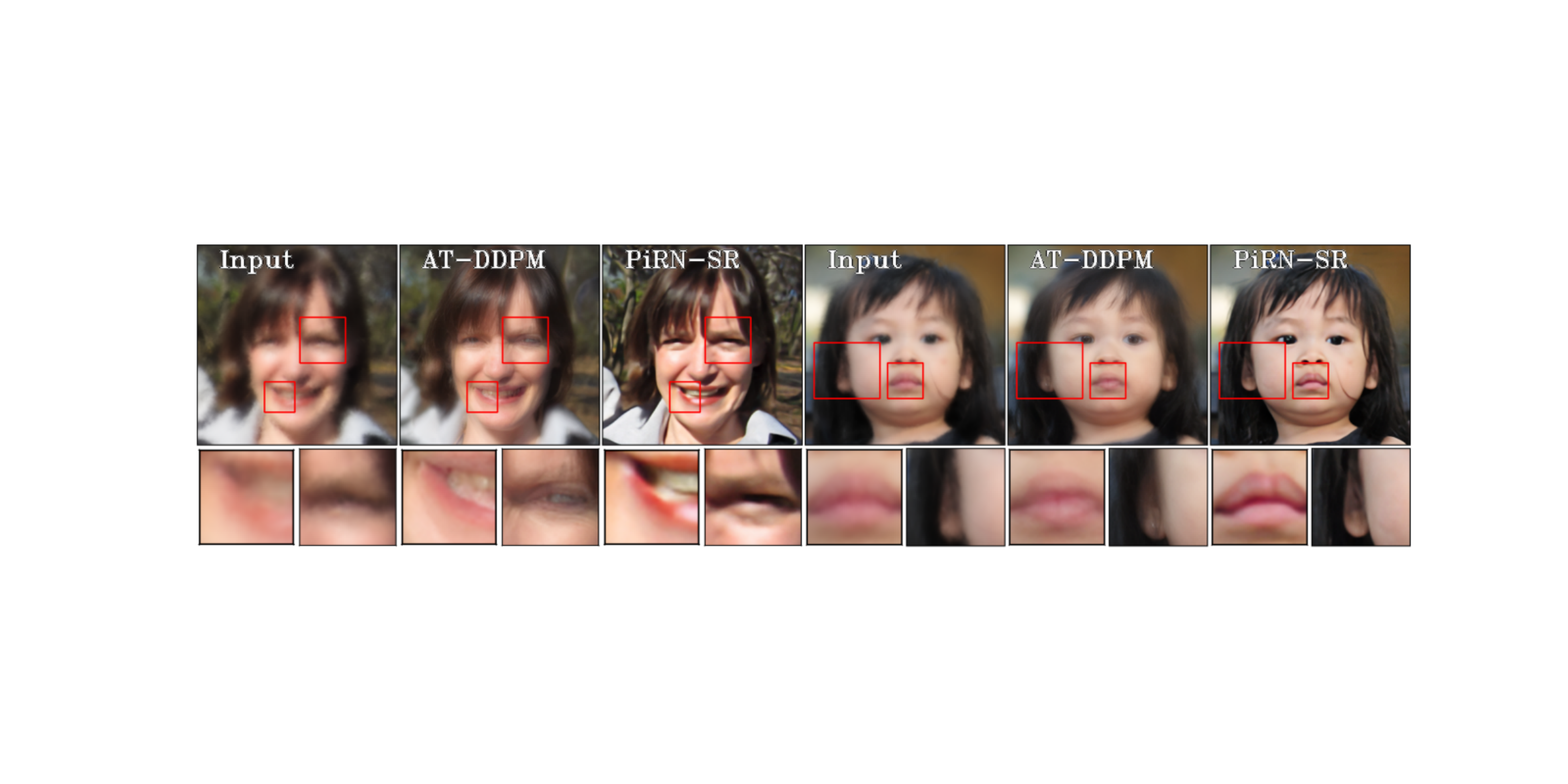}
    \caption{Qualitative performance comparison of AT-DDPM with respect to our proposed framework for atmospheric turbulence mitigation. It can be clearly observed that AT-DDPM hallucinate significantly in comparison with our PiRN-SR output which is assisted by the restoration output of a physics-integrated restoration network (PiRN).}
    \vspace{-0.4cm}
    \label{fig:atddpm_comparison}
\end{figure*}

\subsection{Benefits of PiRN/PiRN-SR for ad-hoc downstream applications}
In this section, we investigate the potential benefits of our proposed restoration network for ad-hoc downstream tasks (RQ4).  We adopted the idea of using the performance of high-level vision tasks, namely text recognition task \cite{Mao_2022_a} and face detection \cite{zhang2016joint}, as an evaluation metric to validate the necessity of restoration.  \textit{Firstly}, for the text recognition task, Figure \ref{fig:real_data_comparison} (Row 1) illustrates the qualitative benefits of our PiRN network with respect to SOTA baseline TurbNet. For quantitative evaluation, we used Average Word Detection Ratio (AWDR) and Average Detected Longest Common Subsequence (AD-LCS) metrics introduced in \cite{Mao_2022_a} for publicly available OCR detection, and recognition algorithms \cite{Shi2015AnET,Tian2016DetectingTI}. Table \ref{table:ocr_challenege} presents the performance gain by PiRN over the real turbulence degraded text images and their restored version by various state-of-the-art methods. OCR algorithms achieve massive improvements of $+0.153$ (AWDR) and +2.422 (AD-LCS) when used on images restored by PiRN compared to being used directly on real images from our proposed test dataset. \textit{Secondly}, for the face detection task, we sampled 15 videos captured from varying distances (200m, 400m, 500m) under turbulence from \cite{cornett2023expanding}. Figure \ref{fig:restoration_detection} presents the performance of ad-hoc MTCNN-based face detector \cite{zhang2016joint} on video frames (distance 400m). It can be clearly observed that our PiRN-SR restored videos significantly help in improving the performance capability of MTCNN ($+0.034$ gain in F1-score). Table \ref{table:bts_mtccn_results} illustrates a detailed F1-score comparison of ad-hoc MTCNN detector (at a confidence threshold $\geq$0.98) on our sampled video subset with and without restoration using recent baselines.

\vspace{-0.05cm}

\subsection{Comparison of PiRN-SR with AT-DDPM} We compare the performance of our proposed approached PiRN-SR with AT-DDPM, which performs knowledge distillation to transfer class prior information from a network trained for image super-resolution to the network for removing turbulence degradation. During inference, rather than starting from pure Gaussian noise, AT-DDPM begins with noised turbulence degraded images for speed-up in inference times. Figure \ref{fig:atddpm_comparison} illustrate the performance comparison of AT-DDPM with respect to our proposed framework for atmospheric turbulence mitigation. It can be clearly observed that AT-DDPM hallucinate significantly in comparison with our PiRN-SR output which is conditioned on high-quality restoration, the output of a physics-integrated restoration network (PiRN). Moreover, instead of completely relying on the diffusion network for complex turbulence restoration, our divide-and-conquer strategy suggests PiRN focus on handling turbulence strength variations and facilitating stochastic refinement (SR) with 10-20 denoising steps to mitigate the gap between the rough restoration and real-world image distribution. Table \ref{table:atddpm_comparison} presents the performance comparison (F1-score) of MTCNN-based face detection AT-DDPM w.r.t. our proposed method and SwinIR. It can be observed that AT-DDPM, as a stochastic method gets lower performance than the deterministic method SwinIR.

\begin{table}[ht]
\centering
\small
\resizebox{\linewidth}{!}{\begin{tabular}{l|ccc|g}
\hline
  &  No Restoration &  SwinIR \cite{liang2021swinir} & TurbNet\cite{Mao_2022_a} & Ours (PiRN-SR)\\
\hline
200m  & 0.5810 & 0.5866  & 0.5923 & \textbf{0.6104}\\
400m  & 0.4987 & 0.4985  & 0.4996 & \textbf{0.5093}\\
500m  & 0.4331 & 0.4006  & 0.4510 & \textbf{0.4648}\\
\hline
\end{tabular}}
\vspace{0.1cm}
\caption{Performance comparison of ad-hoc MTCNN-based face detector \cite{zhang2016joint} on real-world videos captured under atmospheric turbulence \cite{cornett2023expanding} from varying distances. We present the F1-score (at a confidence threshold $\geq$ 0.98, the higher the better) of the ad-hoc MTCNN detector on videos with and without restoration using recent baselines.}
\vspace{-0.2cm}
\label{table:bts_mtccn_results}
\end{table}

\begin{table}[ht]
\centering
\small
\resizebox{\linewidth}{!}{\begin{tabular}{l|ccc|g}
\hline
  &  No Restoration & SwinIR & AT-DDPM Restoration & Ours\\
  \hline
  400m & 0.4987 & 0.4985 & 0.4562 & 0.5093\\
  500m & 0.4331 & 0.4006 & 0.3997 & 0.4648\\
\hline

\hline
\end{tabular}}
\caption{Performance comparison (F1-score) of MTCNN face detector on videos from \cite{cornett2023expanding} with varying distances.}
\vspace{-0.4cm}
\label{table:atddpm_comparison}
\end{table}

\section{Related Work}
\noindent \textbf{Turbulence forward model. }
The simulation of atmospheric imaging can be roughly classified along a spectrum, with pure numerical optics on one side and vision-based simulation on the other. Optics simulations most often come in the form of split-step simulation \cite{Hardie_2017_a, Roggemann_2012_a, Roggemann_1995_a, Schmidt_2010_a}, which numerically propagates waves through a set of random phase screens that represent the atmosphere's spatially varying index of refraction. This method is the most accurate but suffers from a slow speed, preventing the development of large datasets. Computer vision simulations have been utilized \cite{Milanfar_2013_a, Lau_2019_a, Chak_2021_a}, in which pixels are displaced according to some simple local correlations, followed by an invariant Gaussian blur. Despite the fast speed, the domain gap between the simulation and reality is large, limiting its utility for data-dependent restoration or other downstream tasks.

Recently, Zernike-based turbulence simulation methods have been proposed \cite{chimitt2020simulating, mao_p2s, dfp2s}. This method could match the statistics of optics-based simulation and enjoy realistic visual quality while keeping a real-time data synthesis speed. It has been applied to turbulence mitigation \cite{Zhang_2022_a, Mao_2022_a} to facilitate the generalization capability of those models. 

\vspace{0.05cm}
\noindent \textbf{Learning-based turbulence mitigation.}
Image restoration for atmospheric turbulence has been studied since the 1990s \cite{primot1990deconvolution}. Because of the scarcity of data, most methods over the decades are model-based \cite{Delbracio_2015_a, Gilles_2016_a, Huebner_2012_a, Oreifej_2013_a, Anantrasirichai_2013_a, Droege_2012_a, Hardie_2017_b, Milanfar_2013_a, Mao_2020_a}. In recent years, learning-based methods have been proposed, and a detailed introduction is provided in section \ref{sec:intro}. Besides deterministic and stochastic, learning-based turbulence mitigation can also be classified into single-frame \cite{Mao_2022_a, Lau_2021_b, Nair_2021_a, Hoffmire_2021_a, rai2022removing, nair2023ddpm} and multi-frame \cite{Jin_2021_a, Nieuwenhuizen_2019_a, liu2023farsight, Anantrasirichai_2022_a, feng2023turbugan, Zhang_2022_a} ones. 

\vspace{0.05cm}
\noindent \textbf{Diffusion-based stochastic restoration sampler.}
Diffusion or score-based models have been widely used for image restoration tasks as strong posterior estimators. In degradation problems with a known forward model, the unconditional diffusion model can be directly used without any re-training \cite{kawar2022denoising, songsolving, chung2022improving, chung2023diffusion, wang2023zeroshot, song2023pseudoinverseguided}. In those works, forward and reverse models of the degradation functions can be inserted at each denoising step to contract the degraded and clean image manifold. Another line of work aims to solve real-world blind restoration problems without the precise forward model \cite{saharia2022image, saharia2022palette, whang2022deblurring}, it usually requires training a conditional diffusion model with degraded images or latent maps. Diffusion-based restoration sampler could achieve SOTA pixel-wise precision while outperforming deterministic models in perceptual quality by a large margin.

\section{Conclusion}
This paper proposes Physics-integrated Restoration Network (PiRN) to handle stochastic degradation by atmospheric turbulence. PiRN brings the physics-based simulator directly into the training process of a DL restoration paradigm to help the network disentangle the stochasticity from degradation and improve generalization to multiple physical attributes of turbulence. In parallel to improving fidelity, our extended framework PiRN-SR demonstrates how a carefully trained conditional diffusion model can be used as a plug-and-play refiner to generate high perceptual quality restoration results from turbulence-degraded input images at marginal inference overhead. Next, we aim to develop a theoretical understanding to explain the generalization benefits identified by our design.  

\section{Acknowledgments.}
The research is based upon work supported by the Intelligence Advanced Research Projects Activity (IARPA) under Contract No. 2022‐21102100004. The views and conclusions contained herein are those of the authors and should not be interpreted as necessarily representing the official policies, either expressed or implied, of IARPA or the U.S. Government. The U.S. Government is authorized to reproduce and distribute reprints for governmental purposes notwithstanding any copyright annotation therein.

{\small
\bibliographystyle{ieee_fullname}

}

\end{document}